\documentclass[pdflatex,sn-mathphys-num]{sn-jnl}

\usepackage{graphicx}%
\usepackage{multirow}%
\usepackage{amsmath,amssymb,amsfonts}%
\usepackage{amsthm}%
\usepackage{mathrsfs}%
\usepackage[title]{appendix}%
\usepackage{xcolor}%
\usepackage{textcomp}%
\usepackage{manyfoot}%
\usepackage{booktabs}%
\usepackage{algorithm}%
\usepackage{algorithmicx}%
\usepackage{algpseudocode}%
\usepackage{listings}%

\usepackage{caption}
\usepackage{stfloats}
\usepackage{amsmath}
\newcommand{\Tau}{\mathrm{T}}
\usepackage{graphicx}
\usepackage{hyperref}

\theoremstyle{thmstyleone}%
\theoremstyle{thmstyletwo}%

\theoremstyle{thmstylethree}%

\begin{document}

\title[Article Title]{Evaluating the effect of viral news on social media engagement}

\author*[1]{\fnm{Emanuele} \sur{Sangiorgio}}\email{emanuele.sangiorgio@uniroma1.it}
\author[2]{\fnm{Niccolò} \sur{Di Marco}}\email{niccolo.dimarco@uniroma1.it}
\author[2]{\fnm{Gabriele} \sur{Etta}}\email{gabriele.etta@uniroma1.it}
\author[2]{\fnm{Matteo} \sur{Cinelli}}\email{matteo.cinelli@uniroma1.it}
\author[1,3]{\fnm{Roy} \sur{Cerqueti}}\email{roy.cerqueti@uniroma1.it}
\author[2]{\fnm{Walter} \sur{Quattrociocchi}}\email{walter.quattrociocchi@uniroma1.it}

\affil*[1]{\orgdiv{Department of Social Sciences and Economics}, \orgname{Sapienza University of Rome}, \orgaddress{\street{P.le Aldo Moro, 5}, \postcode{00185}, \state{Rome}, \country{Italy}}}

\affil[2]{\orgdiv{Department of Computer Science}, \orgname{Sapienza University of Rome}, \orgaddress{\street{ Viale Regina Elena, 295}, \postcode{00161}, \state{Rome}, \country{Italy}}}

\affil[3]{\orgdiv{GRANEM}, \orgname{Université d'Angers}, \orgaddress{\street{SFR Confluences}, \postcode{F-49000}, \state{Angers}, \country{France}}}


\abstract{This study examines Facebook and YouTube content from over a thousand news outlets in four European languages from 2018 to 2023, using a Bayesian structural time-series model to evaluate the impact of viral posts. Our results show that most viral events do not significantly increase engagement and rarely lead to sustained growth. The virality effect usually depends on the engagement trend preceding the viral post, typically reversing it. When news emerges unexpectedly, viral events enhances users' engagement, reactivating the collective response process. In contrast, when virality manifests after a sustained growth phase, it represents the final burst of that growth process, followed by a decline in attention. Moreover, quick viral effects fade faster, while slower processes lead to more persistent growth. These findings highlight the transient effect of viral events and underscore the importance of consistent, steady attention-building strategies to establish a solid connection with the user base rather than relying on sudden visibility spikes.}

\maketitle
\section{Introduction}
\label{sec:Intro}

The advent and proliferation of social media have fundamentally altered the information landscape \cite{bakshy2012role, schmidt2017anatomy, bergstrom2018news}, offering unprecedented opportunities for content to achieve rapid and widespread attention.
As these platforms have become integrated into our daily lives \cite{grover2022evolution}, transforming into essential tools for information diffusion \cite{pentina2014information, levy2021social} and personal communication \cite{boase2008personal}, they have merged entertainment-driven business models with complex social dynamics \cite{bail2022breaking}, raising significant concerns about their impact on society. This complex interplay produced an environment in which information overload is the foremost feature \cite{bawden2020information} and a wide range of content creators, from news organizations to individual influencers \cite{harrigan2021identifying}, compete for the limited resource that is users' attention \cite{falkinger2008limited, bell2009beyond, anderson2012competition, sangiorgio2024followers}. 

Understanding the attention economy in the digital domain is paramount for navigating this competitive information market, whereby the pursuit of virality \cite{al2019viral} plays a pivotal role in shaping how information sources design their strategies of production and diffusion of content. Characterized by content's exceptional reach and engagement, virality is a core feature of this environment, particularly when referring to viral news and their potential impact on the public discourse. In today's online ecosystem, it is crucial to understand how collective attention responds to abrupt news diffusion and how sudden spikes of visibility reverberate on the subsequent attention captured by the source.
While virality has been mainly investigated for its marketing implications and received extensive coverage in the literature \cite{scott2009new, miller2010social, kaplan2011two, petrescu2011viral, borges2019exploring}, the impact of viral news on collective attention has not obtained as much consideration. 

To address this gap in existing literature, this study aims to enhance our comprehension of the attention economy through a data-driven approach by exploring the dynamics of virality and its effects on users' engagement on different social media platforms. By analyzing data from Facebook and YouTube, we examine attention patterns after viral events to assess how these events influence users' interactions over time. 

In this study, we use a comparative interrupted time-series (CITS) design implemented using a Bayesian structural time series model (BSTS) \cite{scott2014predicting} to evaluate the impact of viral events on users' engagement. In our approach, we apply the BSTS by using increasingly broader time windows to observe the effect of the same viral event from a short-term to a long-term perspective. Based on the BSTS's results, we conduct our analysis first by examining the magnitude of the impact and then its temporal dynamics to address the following two research questions.

\textbf{\textit{RQ1: Does virality induce engagement growth?}} 

Our first research question aims to assess whether and how a viral event leads to increased users' attention received by the source.
After that, our second research question aims to analyze the temporal dynamics of these effects to evaluate if the rapidity at which they occur influences their longevity.

\textbf{\textit{RQ2: Do the faster-manifesting effects persist longer?}}

While our first research question aims to determine the actual impact of virality and its magnitude, the second analysis provides valuable insights into whether these events genuinely contribute to sustained growth or merely act as transient spotlights.
Our results indicate the presence of two different types of viral events, a `loaded-type' and a `sudden-type' virality. When virality follows a sustained growth phase, it represents the final burst of that growth process, with users' attention successively standing on lower levels. Conversely, viral news boosts users' engagement when occurring as a sudden event, reactivating the collective response process. 
While virality can temporarily boost user engagement, this effect is often short-lived. 
We observe that quickly emerging viral effects rapidly fade out. On the other hand, content achieving slower and sustained growth tends to show more persistent effects on engagement. These results emphasize the importance of continuous and consistent content strategies in establishing a solid and enduring connection with the user base rather than relying on viral spikes.

The rest of this paper is organized as follows: Section \ref{sec:literature}, gives an overview of the relevant literature. In Section \ref{sec:Methods}, we outline our methodology for virality detection and impact evaluation. In Section \ref{sec:Results} we present our detailed findings on the dynamics of the virality impact on users' attention and its persistence through time. In Section \ref{sec:Discussion} we discuss the implications of our findings, along with limitations and recommendation for future research. Section \ref{sec:Conclusions} concludes the paper.

\section{Literature review}
\label{sec:literature}

\subsection{Attention economy}
\label{sec:lr_attention}

The attention economy is central to today's digital media landscape. In information management, attention economics applies economic theory to human attention, treating it as a scarce resource. The attention economy is defined as a system of agents (senders) who aim to capture the attention of individuals (receivers) by creating and sharing information packages (signals) \cite{simon1971computers, simon1996designing, davenport2001attention, falkinger2007attention, falkinger2008limited}. Once produced and disseminated, this information undergoes a cognitive filtering process by the receivers, who select relevant information and disregard the rest \cite{treisman1969strategies, kahneman1973attention, zollo2017debunking}.
From a supply-side perspective, the rise of social media platforms has led to an unprecedented volume of available content and information. This overabundance, often referred to as information overload in the literature \cite{jacoby1984perspectives, edmunds2000problem, white2000confronting, bawden2020information}, is a cardinal characteristic of the current attention market in the digital landscape.
In this competitive environment, a wide range of content creators, from news organizations to individual influencers \cite{harrigan2021identifying}, compete for the limited resource that is users' attention. The concept of human attention as a currency entails two key features \cite{Hendricks2019}. Unlike real currencies, it can not be accumulated but solely be spent. Additionally, its transient nature makes it hard to trace and measure. Social media data help address this challenge, as user interactions with content constitute precise engagement indicators. The vast amount of tracked and aggregated data allows for studying large populations, providing insights into collective behavior without the need for experimental settings \cite{wu2007novelty, lazer2009computational}. Shifting the focus from individual users to a broader community perspective highlights the dynamics of collective attention \cite{mocanu2015collective, lorenz2019accelerating}.

\subsection{Human dynamics on social media}
\label{sec:lr_socialmedia}

Originally designed for entertainment and personal connections, social media platforms have become indispensable tools for information dissemination \cite{alipour2024drivers}, raising significant concerns about their potential impact on social dynamics. 
Many works in the literature highlight how online users are prone to consume information aligning with their existing beliefs \cite{cinelli2020selective, del2016spreading, dimarco2024users} and commonly ignore opposing viewpoints \cite{zollo2017debunking, bessi2015science, sultana2023effect}. The creation and reinforcement of online `echo chambers' \cite{del2016echo, terren2021echo}, where shared narratives are collectively shaped and solidified \cite{del2016spreading, nyhan2023like}, may exacerbate social divisions and foster partisan and polarized communities \cite{cinelli2021echo, falkenberg2022growing, allcott2024effects}, complicating the landscape of public discourse \cite{flaxman2016filter} especially during sensitive periods such as global elections \cite{pecile2024mapping}.
Offering unprecedented opportunities for content to achieve rapid and widespread attention \cite{mocanu2015collective}, social media have become crucial environments for the spread of information and misinformation during global events \cite{kim2023details, zhong2023going, shahbazi2024social}, political events \cite{etta2024topology, flamino2023political}, and discussions on emerging technologies such as large language models \cite{alipour2024cross}.
The unprecedented amount of available content distinguishes today’s digital ecosystem, considerably complicating the search for information and giving rise to the phenomenon of infodemics \cite{cinelli2020covid, briand2021infodemics}, such as for the COVID-19 pandemic. These platforms may also influence political landscapes, potentially affecting public opinion and voter behavior during elections through the rapid dissemination and amplification of political content. However, this influence is not definitively proven \cite{bovet2019influence, gonzalez2023asymmetric}.
Research on the interplay between user behavior and platforms' byproduct presented both opportunities and challenges \cite{cinelli2021echo, gonzalez2023social, guess2023social, guess2023reshares}, unveiling a multifaceted landscape where a prevalence of one over the other has not yet emerged. As users' behavior could exhibit persistent patterns across different platforms, topics, and contexts \cite{avalle2024persistent}, the comparative analysis of diverse social media can isolate unaltered consistencies of human dynamics in the digital ecosystem or underline algorithmic and platform peculiarities.

\subsection{Virality on social media}
\label{sec:lr_virality}
On social media platforms, virality refers to a piece of content having rapid diffusion and high levels of users' engagement. 
Viral events can differ depending on the dissemination that leads to their emergence. Structural virality distinguishes between \textit{broadcast} and \textit{viral} diffusion based on their spreading patterns \cite{goel2016structural}. Broadcast diffusion follows a pattern where a single, large parent node spreads information to several smaller entities. In contrast, viral diffusion involves multiple smaller nodes, each contributing to the spread by generating a limited number of infections.

One of the most complex issues is understanding why certain items achieve sudden and widespread dissemination while other similar or higher-quality items remain overlooked \cite{berger2012makes, rathje2023people}. Research suggests that the nature of the content is a more significant factor in driving virality than the characteristics of the spreading source \cite{guerini2011exploring}. For example, the source size in terms of Followers does not affect the probability of going viral on Facebook \cite{sangiorgio2024followers}, with the growth dynamics of engagement exhibiting universal patterns in the short run. Concerning content properties, \cite{bruni2012role} demonstrated that the type of multimedia content on Twitter affects the volume and speed of retweeting, \cite{deza2015understanding} identified visual attributes that can predict relative virality using Reddit data, while \cite{ling2021dissecting} characterized visual elements distinguishing viral from non-viral memes. 

On the other hand, the frantic search for attractiveness may translate into severe drawbacks for the source itself. For instance, \cite{mukherjee2022did} evidenced how readers often view clickbait as a manipulative tactic, which can reduce the perceived competence and trustworthiness of the publisher. 
Information overload leads to similar drawbacks also in business and managerial contexts. The widespread practice of consumer-generated content can provoke social and brand overload due to its quantity and poor quality, leading to negative consequences such as brand disloyalty \cite{lin2023following}. 

Beyond their media features, there is substantial agreement in the literature that emotional resonance is a critical trigger of virality, particularly with negative emotions and extreme or sensitive content. Extreme content, such as sex, nudity, and violence, is more likely to become viral \cite{lance2006subservient}. Hateful content cascades tend to be larger, persist longer, and exhibit more significant structural virality \cite{maarouf2024virality}. An initial burst of a topic's diffusion often correlates with increased negative reactions from users \cite{etta2023characterizing}, and negative messages are typically reposted more quickly and frequently than positive or neutral ones \cite{tsugawa2017relation}.

The spread of misinformation and the rise of political partisanship are critical issues due to their significant social implications. Conspiracy rumor cascades tend to be more persistent and exhibit a positive relationship between their lifetime and size \cite{del2016spreading}. During the COVID-19 pandemic, misinformation was more likely to go viral than truthful information, often due to the use of emotionally charged, other-condemning language \cite{solovev2022moral}. A similar mechanism is observed in political contexts, where posts about political opponents are more likely to be shared on social media. This out-group effect is more influential than other social media-sharing predictors, such as emotional language \cite{rathje2021out}. Ultimately, the driving effects of virality stem from the combination and mutual reinforcement of ideological segregation and negative emotional resonance.

Besides cases of misinformation and harmful content, viral diffusion may also have significant drawbacks in different social contexts. Social media might influence risk perceptions from the general public \cite{vijaykumar2015social}, fueling moral panics and amplifying threats posed by deviant behavior and ideas  \cite{puryear2024moral}, so much so of having tangible consequences in real contexts like inducing distortions and detrimental effects on financial markets \cite{mancini2022self, campbell2023earnings}.

\section{Research design}
\label{sec:Methods}
We now outline our research design, starting from the data collection process detailed in Section \ref{subsec:Data}. In Section \ref{subsec:Defining}, we formally define virality along with the measures to quantify it, while Section \ref{subsec:Detecting} accounts for the virality detection methods. Lastly, in Section \ref{subsec:Evaluating}, we first illustrate our approach to evaluate the virality effect, and then we separately outline our research questions along with their related methods. 

\subsection{Data}
\label{subsec:Data}
We begin by selecting a list of news outlets from NewsGuard \cite{newsguard},  an organization known for monitoring news outlet activities internationally, rating more than 35,000 news and information sources across several countries. We select all the reported news outlets from Germany, France, Italy, and the United Kingdom. NewsGuard provides several categories of descriptive metadata, including the URLs of these news sources' social media accounts, if available.
For Facebook, after identifying all the news outlets with active accounts listed on NewsGuard, we use their Facebook URLs to gather data via CrowdTangle. This Facebook-owned tool monitors interactions on public content from Facebook pages, groups, and verified profiles \cite{crowdtangle}.
Using the `Historical Data' feature in CrowdTangle, we download the complete content history of each page, starting from January 1, 2018. Similarly, for YouTube, after selecting all available accounts, we use the YouTube Data API \cite{youtube} to collect their list of published content. Given the fewer YouTube channels and videos compared to Facebook, we extend the observation period for YouTube to begin on January 1, 2016.

Table \ref{tab:table_dataset} provides an overview of the dataset, detailing the total number of Facebook Pages and YouTube channels used in our analysis for each country, along with their respective number of posts and videos.
The longitudinal structure and temporal continuity of these page observations are crucial for evaluating the impact of viral events. Therefore, the resulting dataset is formatted consistently for both platforms, containing chronological information about each posted and accessible item. This naturally excludes any content that may have been removed by users or through platform moderation at the time of download. Note that in what follows, we will refer to both Facebook pages and YouTube channels as (information) sources and to Facebook posts and YouTube videos as either content or posts, specifying the platform in the latter case.

\begin{table}[t]
\centering
\caption{Dataset overview}
\begin{tabular}{@{}lrrrr@{}}
\toprule
\multicolumn{1}{c}{} & \multicolumn{2}{c}{\textbf{Facebook}}  & \multicolumn{2}{c}{\textbf{YouTube}}   \\ 
\multicolumn{1}{c}{} & \multicolumn{2}{c}{2018-2023} & \multicolumn{2}{c}{2016-2023} \\
\midrule
\textbf{\textit{Country}} & \textbf{\textit{Pages}} & \textbf{\textit{Posts}} & \textbf{\textit{Channels}} & \textbf{\textit{Videos}}\\
Germany & 263 & 9,633,896 & 242 & 193,153 \\ 
France & 252 & 8,705,928 & 177 & 204,665 \\ 
Italy & 384 & 16,259,173 & 234 & 271,183 \\ 
UK & 227 & 8,751,788 & 105 & 100,794 \\ 
\textbf{Total} & \textbf{1,126} & \textbf{43,350,785} & \textbf{758} & \textbf{769,795} \\ 
\bottomrule
\end{tabular}
\label{tab:table_dataset}
\end{table}


\subsection{Defining virality}\label{subsec:Defining}
On social media platforms, virality represents the propensity of content to achieve rapid diffusion and high interactions levels from users \cite{al2019viral}. To quantify virality, we use a bivariate approach based on two measures, spreading and interactions. The following subsections provide the definition of these two measures for each social media platform.

\subsubsection{Spreading}
For Facebook data, we adopt the metric provided by \cite{elmas2023measuring}. Originally defined on Twitter data, the proposed metric can be applied to any social media platform with equivalent Retweets and Followers metrics. On Facebook, Shares are equivalent to Retweets, representing the number of users sharing certain content. On both platforms, Followers represent the number of users subscribed to a given page at the time of posting. Therefore, on Facebook, we can define the Spreading $S$ of a given post $j$ of a Page $i$ as:
    \begin{equation}
    S_{ijt} = ln\left(\frac{Shares_{ij}}{Followers_{it}}\right)
    \end{equation}
    where $Followers_{it}$ indicates the number of Followers of the Page $i$ at time $t$, i.e. the time of posting.

For YouTube, we can define it using the video Views metric, which specifically measures content spreading. Therefore, on YouTube, we define the Spreading $S$ of a given video $j$ published by a Channel $i$ at time $t$ as:
    \begin{equation}
    S_{ijt} = ln\left(Views_{ij}\right) .
    \end{equation}

\subsubsection{Interactions} 
On Facebook, interactions are measured by the Total Interactions, encompassing the total number of users' interactions with the post (the sum of Likes, Comments, and Shares). Therefore, on Facebook, we define the Total Interactions $TI$ of a given post $j$ of a Page $i$ as:
    \begin{equation}\label{eq:TI_fb}
    TI_{ijt} = ln\left(Likes_{ij} + Comments_{ij} + Shares_{ij}\right) 
    \end{equation}
    where $t$ is the date in which post $j$ appears.

Since YouTube does not provide a share count (and consequently its Shares metric), we define the Total Interactions $TI$ of a given video $j$ of a Channel $i$ published at day $t$ as:
    \begin{equation}\label{eq:TI_yt}
    TI_{ijt} = ln\left( Likes_{ij} + Comments_{ij}\right) .
    \end{equation}

\subsection{Detecting virality}\label{subsec:Detecting}
We measure virality on social media as an (exceptional) over-performance of content regarding spreading and users' interactions. To detect virality, we first apply a $z-$score normalization to both $S_{ijt}$ and $TI_{ijt}$ for each piece of content $j$ from a source $i$. On Facebook, a post is considered {\it viral} if both its $z-$scores exceed 3. We set the threshold for YouTube at 2.5 to ensure a comparable number of videos and channels as in Facebook.

\begin{figure*}[t]
       \centering
    \includegraphics[width=0.9\textwidth]{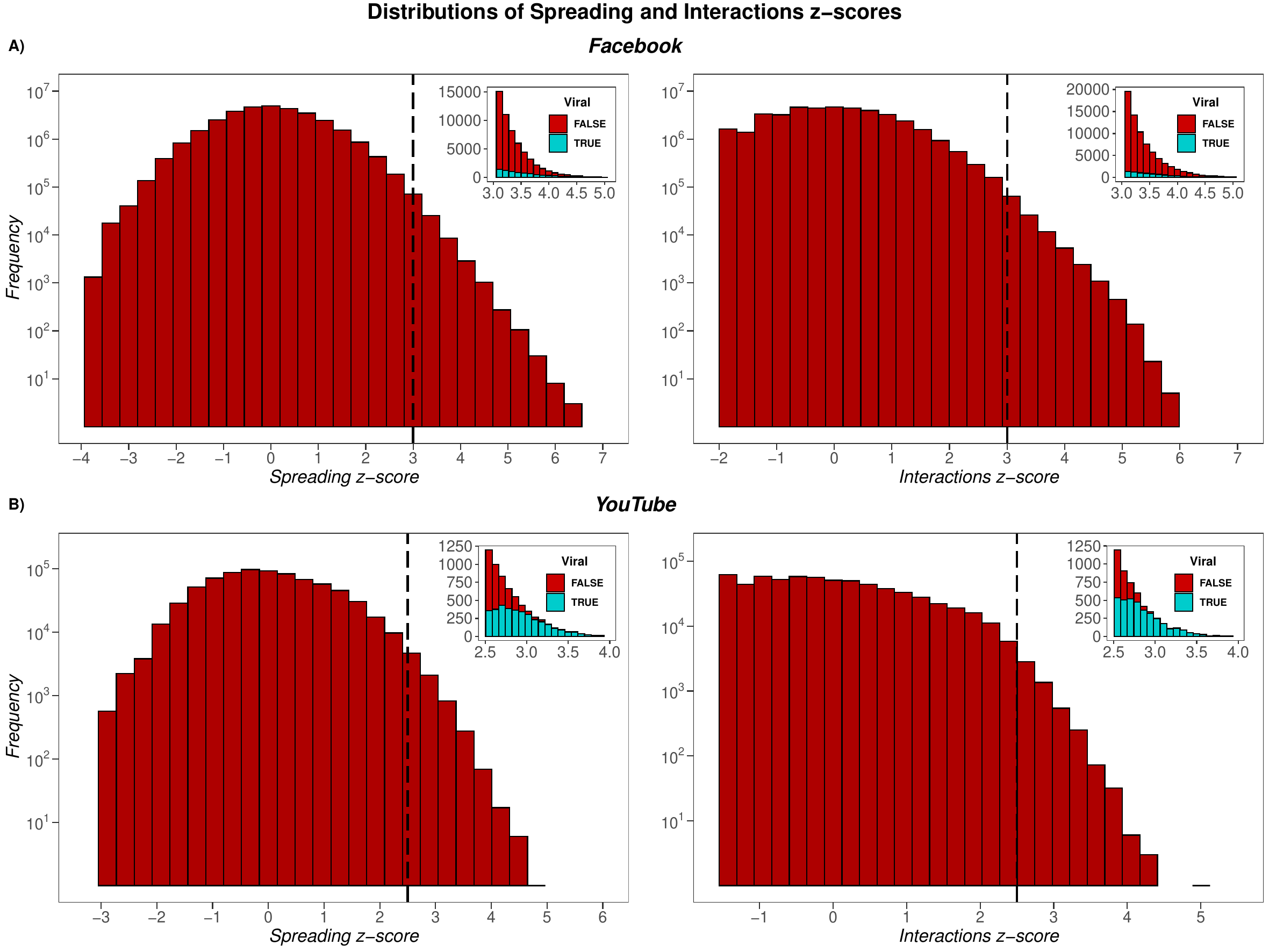}
        \caption{Distributions of Spreading and Engagement $z-$scores for Facebook and YouTube. The inset plots of each chart display the subset of values exceeding the threshold for the given metric, showing its breakdown between viral contents (which exceed the threshold in both metrics) and non-viral ones (hence exceeding the threshold only in the displayed metric).}
        \label{fig:zscores_distributions}
\end{figure*} 

The distributions of the two $z-$scores for both platforms are shown in Fig. \ref{fig:zscores_distributions}. The inset plots within each panel zoom on the subset of values that exceed the threshold for the displayed metric, distinguishing between viral content (which surpasses both thresholds) and non-viral content (which exceeds the threshold only in the displayed one).  

These insets reveal that the subset resulting from the combination of overperformances constitutes a small fraction of the distribution tails, especially on Facebook. In other words, content may induce high interactions levels without widespread diffusion, or conversely, it may fail to engage users despite extensive reach. This observation further emphasizes the exceptional rarity of virality. 

\begin{figure*}[t]
       \centering
    \includegraphics[width=0.7\textwidth]{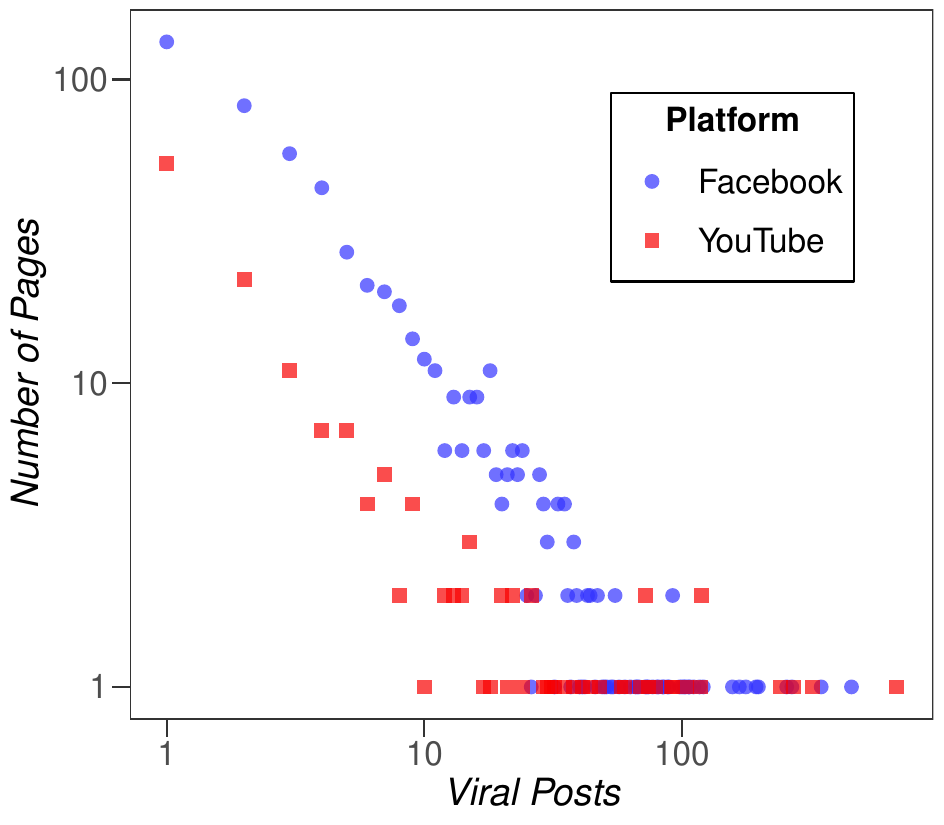}
        \caption{Distributions of viral posts per source on Facebook and YouTube.
  Facebook sample sizes: Pages = 613, Posts = 9514; 
  YouTube sample sizes: Channels = 161, Videos = 3624. }
        \label{fig:viral_posts}
\end{figure*}

Fig. \ref{fig:viral_posts} presents the distributions of viral posts per source for Facebook and YouTube. As the Figure shows, despite the differences in sample sizes, both platforms exhibit similar heavy-tailed distributions over different scales reflecting the peculiar dynamics of content virality. If, on the one hand, a limited number of sources succeeded in achieving virality multiple times, it represented a unique or rare event for most of them.

\subsection{Evaluating virality impact}\label{subsec:Evaluating}

After defining the procedure for detecting viral posts, we can now delve deep into our approach for evaluating the impact of viral events on the performance of Facebook pages and YouTube channels.

In terms of attention received, we quantify the performance of a given source by calculating its daily engagement. We define the {\it Engagement} $E_{it}$ of a source $i$ at day $t$ as the sum of the Total Interactions $TI$ (as in Eq. \ref{eq:TI_fb}, and \ref{eq:TI_yt}) of all its $j$ posts published that day, i.e.:
$$
E_{it}=\sum_{j}TI_{ijt}.
$$
In assessing the total attention a news outlet receives, we interpret higher engagement as indicative of greater user attention, regardless of the number of posts published. This approach is based on the premise that increasing the content volume only results in heightened engagement if the content effectively captures users' attention. Conversely, high engagement across several posts implies high users' attention, and calculating its average value would potentially underestimate this effect \cite{sangiorgio2024followers}.

Our analysis aims to evaluate the after-effects of a viral event on the source's performance in terms of attention caught. To inspect the attention dynamics following the event, we first assess whether the engagement received significantly changes after it. 
To detect a significant variation (either positive or negative) on the engagement after a viral event, we use a comparative interrupted time-series (CITS) design implemented using a Bayesian structural time series model (BSTS, hereafter) \cite{scott2014predicting}, which has also been used in previous research \cite{trujillo2022make}. In our approach, we apply BSTS to each viral event, using time windows associated to $n$ weeks ahead and before, with $n=2,3,4,5,6$, constructed as follows: for any given $n$, we exclude the day of the viral post and compare the engagement trend during the $n$ weeks following the event to the expected trend based on the $n$ weeks preceding it -- hence, having a windows of $2n \times 7$ days -- and controlling for the presence of other viral posts. Based on the BSTS's results, we then conduct our analysis to address our two research questions outlined in the following sections, first examining the magnitude of the impact and then its temporal dynamics.

\subsubsection{\textbf{Research Question 1:} Does virality induce engagement growth? }
\label{subsec:RQ1}
\vspace{2pt}

Our first research question aims to assess whether - and how - a viral event increases users' attention received by the source after its occurrence. We start from the premise that each viral post can have a positive, null, or negative effect on the subsequent engagement in the considered time window. 

To evaluate the effect, we use two variables provided by the BSTS: 1) the Average Absolute Effect, which indicates whether the impact was positive or negative, along with its magnitude, and 2) the statistical significance of the effect, which is captured by the $p$-value $p_\alpha$ of BSTS at a confidence level $\alpha$. For the analysis, we set $\alpha=0.05$.

If the effect has no statistical significance at a confidence level $\alpha$, we consider that virality did not have a discernible impact on users' engagement - which we refer to as \textit{No Effect}. Otherwise, if the effect is statistically significant and the observed engagement significantly deviates from the expected trend based on the examined $n$ weeks, we consider that virality had a \textit{Growth} or \textit{Decrease} effect on users' attention as quantified by the Average Absolute Effect. For our analyses, we use the BSTS implementation from the CausalImpact R package \cite{brodersen2015inferring}. This research question is addressed in Sec. \ref{sec:RQ1}.




\subsubsection{\textbf{Research Question 2:} Do the faster-manifesting effects persist longer?}

\vspace{2pt}
To continue our investigation of virality dynamics, we aim to examine the relationship between the speed at which effects manifest and their persistency over time. By using increasingly wider time windows, we can observe the effect of the same viral event from a short-term to a long-term perspective. 

For each detected effect, its emergence and its subsequent persistency can be determined through a classification based on the output from the BSTS. For each viral post $j$ of a page $i$, we define the time of {\it emergence} of its effect, denoted as $\tau^{(em)}_{ij}$, as follows:
\begin{equation}\label{eq:t_em}
\tau^{(em)}_{ij}=\min\{n=2,3,4,5 : p_{ij}(n)<p_\alpha\},
\end{equation}
where $p_{ij}(n)$ is the $p$-value of BSTS applied to the time window associated to week $n$.
Similarly, the {\it fade-out} time, $\tau^{(f-o)}_{ij}$, is identified as the earliest time window in which the previously detected effect no longer manifests, i.e.,
\begin{equation}\label{eq:t_fo}
\tau^{(f-o)}_{ij}=\min\{n> \tau^{(em)}_{ij} : p_{ij}(n)>p_\alpha\}.
\end{equation}
We notice that the sets associated to the definitions of $\tau^{(em)}_{ij}$ and $\tau^{(f-o)}_{ij}$ are not empty in the cases treated in this paper.

Notice also that the $\tau$'s lead to a partition of the collection of the posts with the related pages. Specifically, fixed $\bar{n}$, we define

\begin{equation}
   \Tau_{\bar{n}}^{(em)}=\{(i,j) \in \mathcal{I} \times \mathcal{J} : \tau^{(em)}_{ij}=\bar{n}\} 
\end{equation}

and 
\begin{equation}
\Tau_{\bar{n}}^{(f-o)}=\{(i,j) \in \mathcal{I} \times \mathcal{J} : \tau^{(f-o)}_{ij}=\bar{n}\}.
\end{equation}
We define the {\it persistency} $\phi_{kh}$ as the fraction of posts still having persistent effects at a given week $h$ after their emergence time $k$, i.e.,
\begin{equation}\label{eq:per}
\phi_{kh}=\frac{|\Tau_{k}^{(em)}|-\sum_{\ell=k+1}^h|\Tau_{\ell}^{(f-o)}|}{|\Tau_{k}^{(em)}|},
\end{equation}
where $|\bullet|$ is the cardinality of set $\bullet $, for each $k$ {\(\displaystyle \in [2, 5]\)} and $h$ {\(\displaystyle \in [k+1, 6]\)}. See Sec. \ref{sec:RQ2} for further details, in which we address the Research Question 2. 
  
By analyzing the persistency rates across all cases for each specified emergence time, we can evaluate the relationship between the speed of manifestation and the longevity of the effects. This analysis will help determine whether effects that emerge more quickly tend to persist longer or, conversely, if they vanish more rapidly than slow-emerging ones.

This approach helps us to understand the temporal behavior of viral effects, as we can further extrapolate the observed dynamics to their unobservable earlier phase by tracing back to the time of the event and evaluating the dynamics as we approach it asymptotically. 
This analysis provides valuable insights into how collective attention responds to sudden stress conditions, like viral news events, and to determine whether these events genuinely contribute to sustained growth or merely act as transient spotlights.

\section{Results}
\label{sec:Results}

\begin{figure*}[t]
       \centering
    \includegraphics[width=\textwidth]{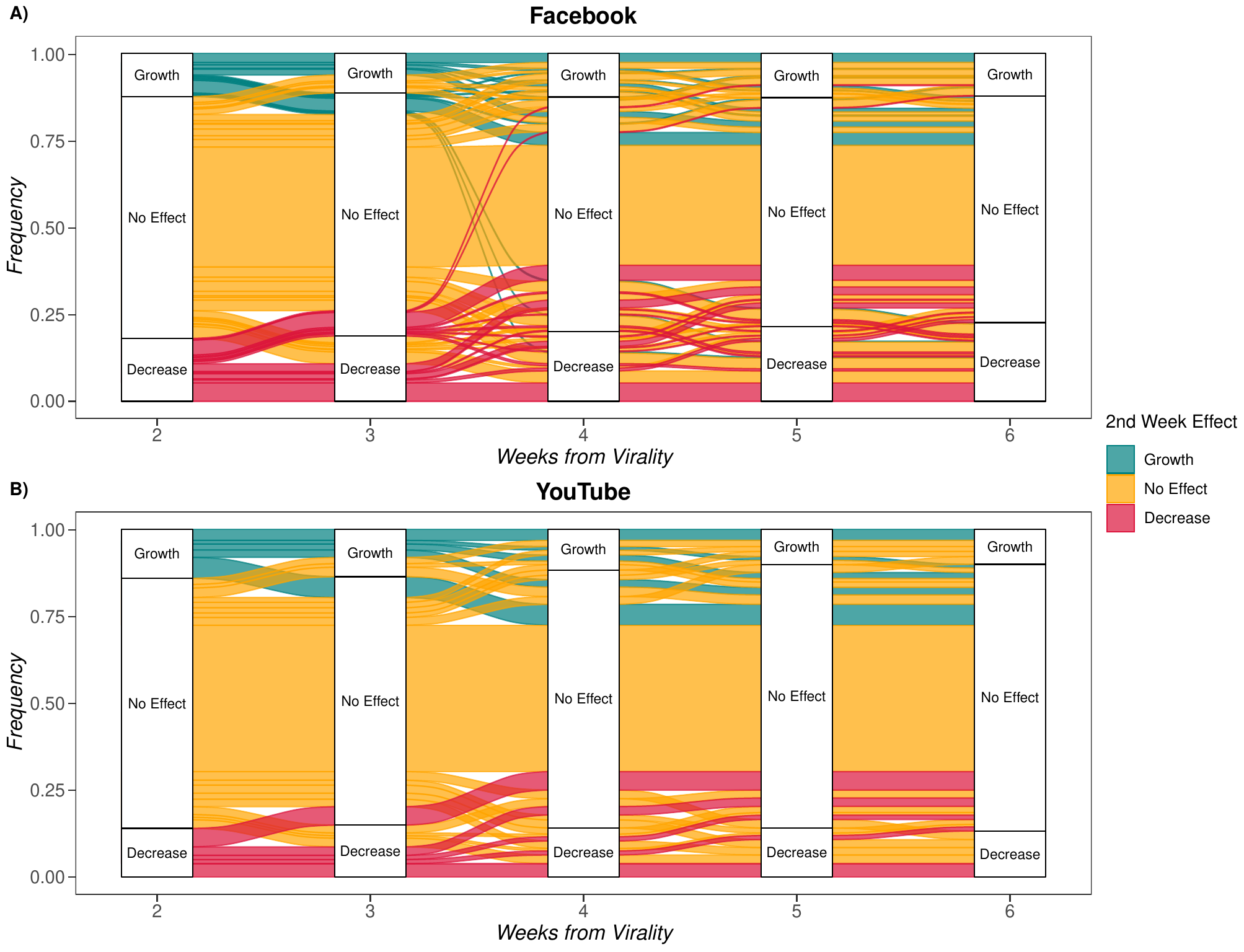}
        \caption{Effects after virality with increasing time windows. The $y$-axis represents the percentage of Growth, No Effect, and Decrease cases, for each time scale on the $x$-axis. Links represent the effects flow between consecutive weeks, color-coded according to the first observed effect (i.e., in the 2-week window).}
        \label{fig:effect_weeks}
\end{figure*} 

Before discussing our research questions in depth, we first inspect the BSTS's results, which provide us with a useful initial overview to help delve into the analysis. We begin by examining the proportion of cases that exhibit positive, null, or negative impacts and assess their consistency across various time windows. 

The results are reported in Fig \ref{fig:effect_weeks}, which shows the percentage of Growth, No Effect, and Decrease cases on the $y$-axis for each time scale on the $x$-axis. The diagram includes links representing the effects flow between consecutive weeks, color-coded according to the first observed effect to maintain a clear visual trajectory through the data stream. A significant observation from this analysis regards the high percentage of cases showing no statistically significant impact on engagement performance. This first finding suggests that virality does not necessarily enhance engagement, as it often results in an indiscernible impact.

\begin{figure*}[t]
       \centering
    \includegraphics[width=\textwidth]{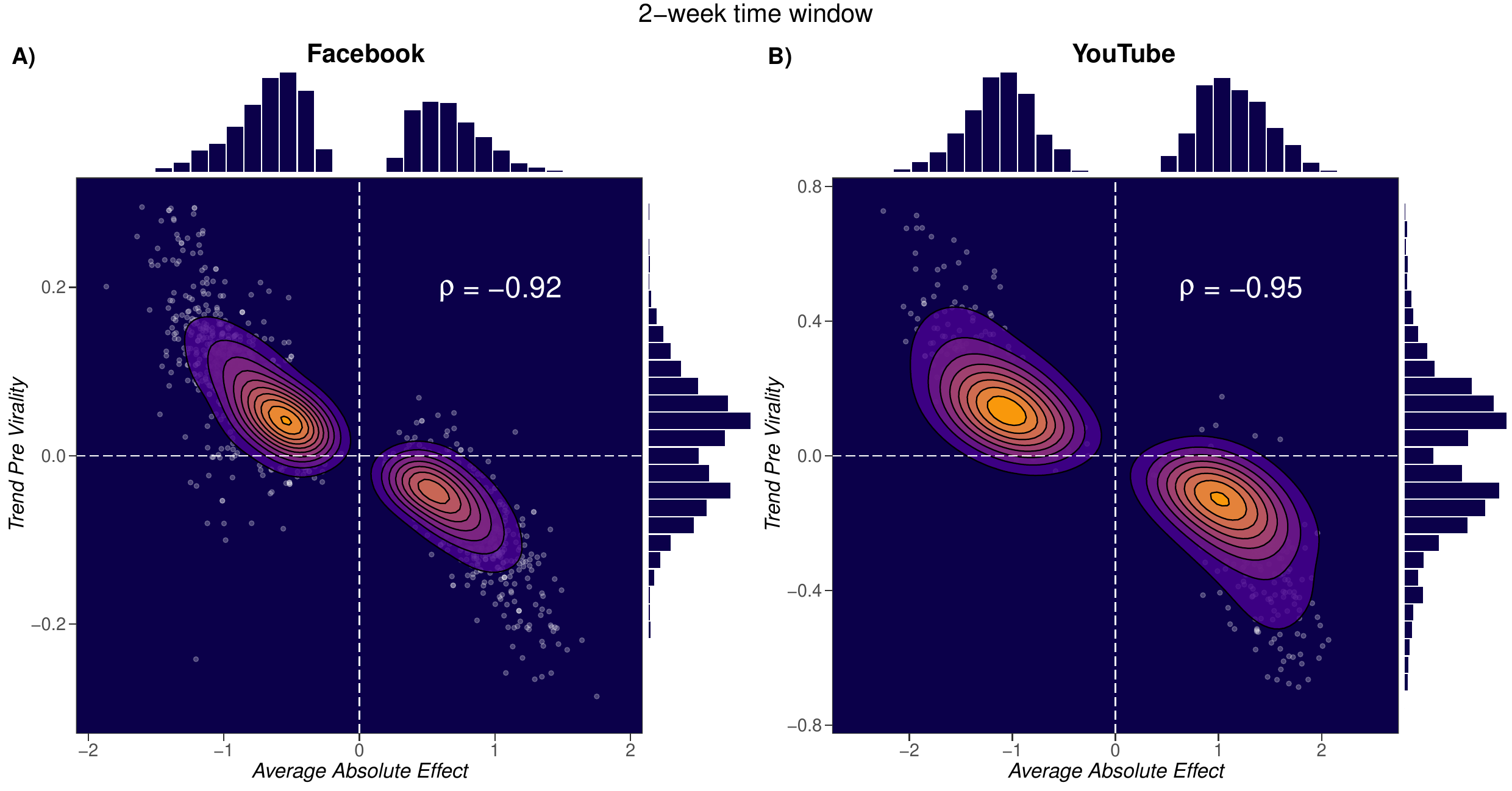}
        \caption{Density of the trend preceding the viral post and the average absolute effect on engagement for the 2-week time window. Trend Pre Virality is the $\beta1$ coefficient of the regression estimated by the BSTS on the weeks preceding virality. Given its previous trend, the average Absolute Effect is the average effect on the Engagement after the viral event. Only events with a statistically significant effect on Engagement are shown.}
        \label{fig:Density_effect}
\end{figure*} 

Secondly, the percentages of Growth and Decrease cases are comparable, with slightly higher occurrences of negative impacts. This suggests that, apart from being infrequent, the impact of a viral event on users' attention can even be detrimental. Contrary to common expectations \cite{al2019viral,tellis2019drives,malodia2022meme}, virality rarely induces engagement growth.
Furthermore, we observe that effects typically emerge or shift between the second and fourth weeks, with the most significant transitions happening from the third to the fourth week. Beyond this period, the consistency of effects is broadly stable. Given the inherently short-term nature of virality, effects observed in larger time scales, such as 5- and 6-week windows, are less likely to be directly attributable to the viral event. These longer time windows are primarily aimed at assessing the persistency of earlier effects, enabling us to analyze the dynamics from a short-term to a long-term perspective and to distinguish the boundary between these periods. 

\subsection{RQ1: Dynamics of virality effect on Engagement}
\label{sec:RQ1}
We now deepen the dynamics of the virality effect on Engagement by focusing exclusively on the Growth and Decrease cases, as defined in Sec. \ref{subsec:RQ1}. 

In Fig. \ref{fig:Density_effect}, we present the joint density of the slopes of the attention trends preceding the viral posts and the absolute effects on engagement after the event. 

The `Trend Pre Virality' on the y-axis represents the $\beta_1$ coefficient of the regression estimated by the BSTS for the $n$ weeks preceding the viral event. On the x-axis, we show the Average Absolute Effect on Engagement in the examined $n$-week window after virality, accounting for its previous trend.

\begin{figure*}[t]
       \centering
    \includegraphics[width=0.9\textwidth]{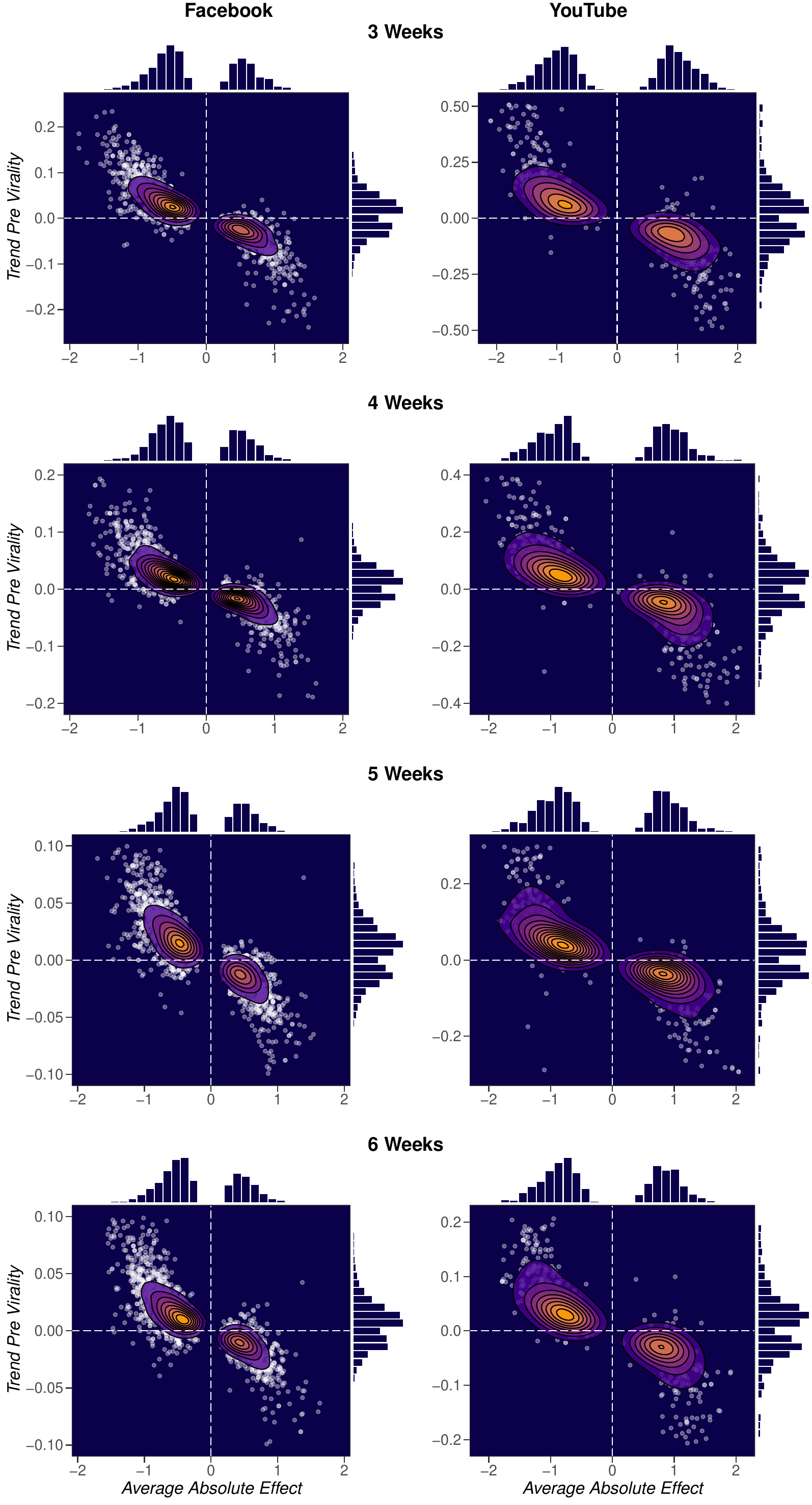}
        \caption{Density of the trend preceding the viral post and the average absolute effect on engagement for the 3, 4, 5, and 6-week timescales. We recall that Trend Pre Virality is the $\beta1$ coefficient of the regression estimated by the BSTS on the weeks preceding virality, and that the Average Absolute Effect is the average effect on the Engagement after virality estimated by the BSTS, given its previous trend. Only events with a statistically significant effect on engagement are shown.}
        \label{fig:Density_other}
\end{figure*}

Fig. \ref{fig:Density_effect} reports the values for the 2-week window as an example, while Fig. \ref{fig:Density_other} in SI Appendix shows the results for other timescales on both platforms which display consistent results.

As Fig. \ref{fig:Density_effect} shows, the density is split into two opposing quadrants: from growth to decrease and from decrease to growth, highlighting a significant negative correlation between the preceding trend and the consequent absolute effect. This dynamic shows consistency on both platforms and across their timescales, as shown in Tab \ref{tab:tab_correlation}, which reports the Spearman’s correlation coefficients between the $\beta_1$ coefficient of the trend preceding virality and the Average Absolute Effect on Engagement. 

\begin{table}[ht]
\captionsetup{font=small}
\centering
\caption{Spearman's correlation coefficients between the $\beta1$ coefficient of the trend preceding virality and the Absolute Effect on Engagement for different platforms and time windows. Posts and Videos represent the number of events for which a Growth or Decrease effect is detected, according to the BSTS.}

\begin{tabular}{@{}ccccc@{}}
\toprule
\multicolumn{1}{c}{} & \multicolumn{2}{c}{\textbf{Facebook}}  & \multicolumn{2}{c}{\textbf{YouTube}}   \\ 
\midrule
\vspace{2pt}
\textbf{\textit{Timescale}} & $\boldsymbol{\rho}$ & \textbf{\textit{Posts}} & $\boldsymbol{\rho}$ & \textbf{\textit{Videos}} \\ 
2 Weeks & -0.92 & 3015 & -0.95 & 1100 \\ 
3 Weeks & -0.91 & 2953 & -0.93 & 1048 \\ 
4 Weeks & -0.90 & 3134 & -0.93 & 1049 \\ 
5 Weeks & -0.88 & 3279 & -0.93 & 1028 \\ 
6 Weeks & -0.87 & 3399 & -0.93 & 1016 \\ 
   \hline
\end{tabular}
\label{tab:tab_correlation}
\end{table}

Therefore, virality positively impacts users' engagement when occurring as a sudden event on a declining collective attention. Conversely, when virality manifests following a sustained growth phase, it represents the final burst of that growth process, with users' attention successively standing on lower levels than its previous phase. 

These results shed light on the bounded yet elastic nature of collective attention. While additional growth following a sustained growth phase is extremely rare, viral events act as a booster when users' attention is likely to nearing its lower bound, reactivating the collective response process.

By focusing on viral news, these results potentially indicate the presence of two different types of viral events. The first is a `loaded-type' virality, where attention progressively increases, culminating in the viral event. This type could occur in scenarios where information is gradually revealed, such as sequences of rumors, confirmations, and official announcements. The second type represents a `sudden-type' virality, with the news emerging unexpectedly as an exogenous event. Based on the release patterns of information, this interpretation could explain the differences between the two types, along with the inverse relationship between the preceding attention pattern and the after-effect of the viral news.

\subsection{RQ2: Emergence and persistency of the virality effect}
\label{sec:RQ2}

After assessing how virality affects users' engagement and its extent, we now inspect the temporal dynamics of the effect by examining the relationship between its manifestation and longevity. In this section, the analysis will specifically focus on the effect's persistency based on its timing of emergence and fade-out.

In Eq. \ref{eq:t_em}, we defined the time of $\tau^{(em)}_{ij}$ as the shortest time window during which a Growth or Decrease effect first becomes significant. For example, if the impact of a viral event is significant within a 2-week time window, we assign its emergence time as 2. Conversely, if the impact is not significant within the 2-week window but becomes significant in the 3-week window, the emergence time is then 3.

Similarly, we examine the duration for which the effect remains unchanged and determine the window in which it ceases to be consistent. As we defined in Eq. \ref{eq:t_fo}, the {\it fade-out} $\tau^{(f-o)}_{ij}$ of the effect corresponds to the first window where it no longer manifests. For example, if an event consistently exhibits a Growth effect from the 2-week to the 4-week window and then ceases to manifest in the 5-week window, the {\it emergence} time would be 2, and the {\it fade-out} time would be 5.

Consequently, as we defined in Eq. \ref{eq:per}, the {\it persistency} $\phi_{kh}$ is the fraction of still persistent effects at $h$ weeks after their given emergence time $k$, to which we generally refer here. 
Hence, we evaluate the effect's decay by calculating the fraction of still persistent effects after each week, grouped by their time of emergence. For instance, over the total number of cases where the impact appears two weeks after virality, we calculate the proportion of persistent effects in the 3rd, 4th, 5th, and 6th week. This method allows us to derive the effect's decay curve for each time of emergence $k$ {\(\displaystyle \in [2, 5]\)}.

At this stage, we are able to observe the persistency rate up to a minimum of two weeks after the viral event. By fitting a negative exponential function of $k$ and power law in $h$, we can estimate the variation of the exponent of the decay curves based on their time of emergence $k$ as:

\begin{equation}
\resizebox{!}{0.48cm}{$ \phi_{kh} = f(k,h) = h^{-\lambda+\frac{k}{\beta}} $}.
\end{equation}

This procedure allows us to describe the decay behavior of the effects based on their emergence and to extend it to their unobservable earlier phases as if we approach them asymptotically.

\begin{figure*}[b]
       \centering
    \includegraphics[width=0.95\textwidth]{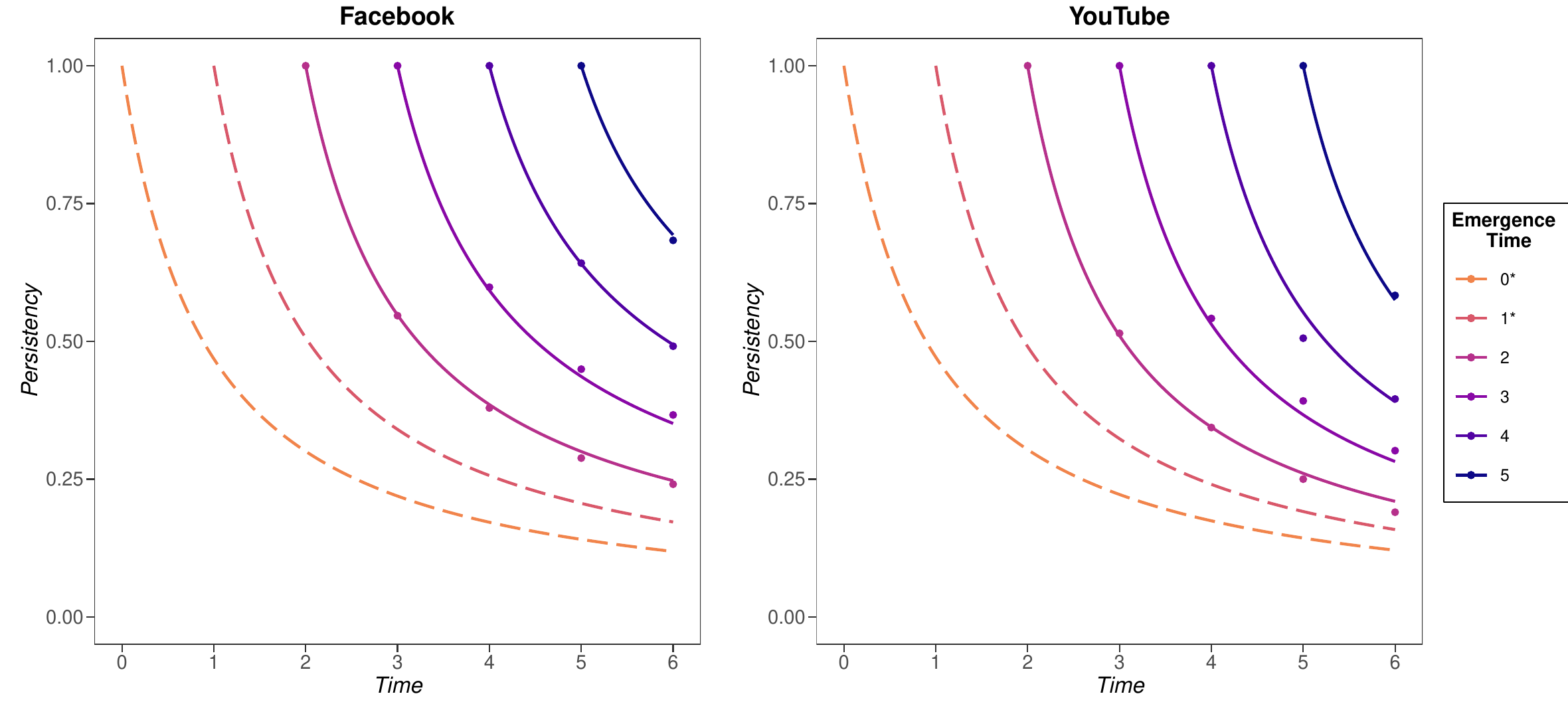}
        \caption{Persistency of the virality effect based on its time of emergence. Solid lines represent the estimated decay for the observed curves - from the 2nd week to the 5th week of emergence time - along with their observed values. The dotted lines represent the corresponding extrapolated decay curves for 0 and 1 week after virality as emergence time.}
        \label{fig:persistency}
\end{figure*}

The results of the fitting procedure are reported in Tab. \ref{tab:tab_param}, while Fig. \ref{fig:persistency} shows the graphical representation of the decay curves for each emergence time. Solid lines represent the estimated decay for the observed curves - from the 2nd week to the 5th week of emergence time - along with their observed persistency. The dotted lines represent the corresponding extrapolated decay curves for 0 and 1 week after virality as emergence time. The curves representing extrapolations inherently involve higher uncertainty and should be considered useful asymptotic approximations for understanding the dynamics during their unobservable phases. 

\begin{table}[ht]
\captionsetup{font=small}
\caption{Paramaters estimation}
\centering
\begin{tabular}{lcrrrr}
\toprule
Platform & Parameter & Estimate & Std. error & p-value  
\vspace{2pt}\\ 
\midrule
Facebook & $\lambda$ & 1.09 & 0.02 &  \textless 0.001 \\  
Facebook & $\beta$ & 8.84 & 0.44 &  \textless 0.001 \\  
YouTube & $\lambda$ & 1.09 & 0.05 &  \textless 0.001 \\ 
YouTube & $\beta$ & 17.44 & 4.55 & 0.0024 \\  
\bottomrule
\end{tabular}
\label{tab:tab_param}
\end{table}

According to the estimations, in about 50\% of cases, the impact either fades out within the first week following the viral event or does not occur.
Moreover, the data consistently show that earlier emergences of viral impacts are associated with faster decay across both platforms. This indicates that faster processes tend to fade quicker, while slower ones exhibit more persistence. This finding highlights the elastic nature of collective attention when stretched to its limits. The volatile and fluctuating nature of attention prevents it from being steadily focused, resulting in a trade-off between the rapidity of the effect and its durability. These observations have critical implications for content producers in the digital realm, underscoring a clear distinction between short-term and long-term dynamics of collective attention.

From a probabilistic perspective, sudden and disproportionate growth is rare and rarely leads to a noticeable positive impact on engagement. Even when it occurs, its effects tend to fade away swiftly. Conversely, organic and sustained growth, though slower to manifest, tends to have more enduring effects. This contrast emphasizes the transient nature of viral events compared to the lasting effectiveness of consistent engagement, highlighting the importance of gradual and continuous attention-building strategies rather than relying on abrupt surges of visibility.

\section{Discussion}\label{sec:Discussion}

\subsection{Implications}
Our findings highlight collective attention's bounded and elastic nature under sudden stress conditions, such as viral events, and have significant implications for the information sources ecosystem in the digital domain. Firstly, our analysis reveals that viral events rarely lead to engagement growth, suggesting that the frantic pursuit of sudden visibility is often unproductive. The study distinguishes between two primary types of virality, each corresponding to a different mechanism of collective attention response. The first type, `loaded-type virality,' is characterized by a gradual increase of engagement that culminates in a final burst of attention in the viral event. The second type, `sudden-type virality,' occurs when news emerges unexpectedly, similar to an exogenous event, and is the only scenario in which virality leads to an increased users' response.
Regardless of whether the impact is positive or negative, our findings indicate that effects emerging more quickly tend to fade faster, while slower-emerging processes are more persistent over time. As well as being an extremely rare event, virality does not turn out to be an effective long-term growth strategy. These insights underscore the advantages of organic and continued growth strategies in establishing a solid and enduring connection with the user base, yielding positive and sustained responses regarding collective attention.

\subsection{Limitations and Future Research}
The findings of this study are specifically tied to the platforms analyzed, the type of pages examined, and the period considered. Although the results are consistent with general human dynamics observed in other collective attention studies, they cannot be generalized across all social media platforms. Each platform has distinct user bases and algorithmic characteristics that influence how content is spread and engages users.
Additionally, the specificity of the sources analyzed — news outlets - and their content type must be considered. Focusing exclusively on news items offers the advantage of dealing with content inherently tied to specific topics or events, providing a concrete subject matter. However, the dynamics and peculiarities of the news agenda mean that these results may not necessarily apply to the broader spectrum of content creators.
Future research should explore various social media platforms and content types to understand better the dynamics of virality and collective attention across different contexts. A more comprehensive picture of how virality functions in the digital landscape can be developed by expanding the study to include a broader range of content creators and user engagement patterns.

\section{Conclusions}
\label{sec:Conclusions}
This study offers a detailed analysis of virality dynamics on social media platforms. It examines how viral events affect users' engagement and the relationship between their rapid emergence and subsequent persistency. Our findings challenge the common assumption that virality regularly enhances user engagement.
While viral events may momentarily capture attention, our evaluation reveals that they rarely foster sustained engagement growth. Indeed, the impact of virality typically depends on the preceding growth trend, with a pronounced negative correlation observed between the two.
We categorize viral events into two distinct types: `loaded-type' virality, characterized by a gradual build-up of attention culminating in a viral burst, and `sudden-type' virality, which appears unexpectedly like an exogenous shock. 

When virality occurs following an ongoing growth phase, it represents its final peak, resulting in diminished attention levels. In contrast, a viral event reactivates the collective response process when arising suddenly after a declining attention phase.

Additionally, our study highlights the elastic nature of collective attention, demonstrating that effects emerging swiftly tend to fade quickly, whereas slower-emerging effects show greater persistency. This finding highlights the critical role of content producers in fostering consistent, high-quality engagement rather than depending solely on transient viral spikes. The rapid dissipation of viral impacts, particularly those that emerge quickly, illustrates the volatility and fluctuation of collective attention. This volatility implies a trade-off between the rapidity of an impact's emergence and its lasting presence, emphasizing content producers' challenges in capturing and maintaining users' engagement.
In conclusion, although virality can temporarily surge users' attention, its effects are usually short-lived, making the frantic pursuit of sudden visibility an often fruitless strategy.

\section{Author contributions statement}
\begin{itemize}
    \item \textbf{Emanuele Sangiorgio}  Conceptualization, Methodology, Validation, Data collection, Data curation, Data analysis, Visualization, Writing.
    \item \textbf{Niccolò Di Marco} Methodology, Validation, Data analysis, Writing.
    \item \textbf{Gabriele Etta} Data collection, Data curation, Data analysis, Writing.
    \item \textbf{Matteo Cinelli} Conceptualization, Methodology, Supervision, Writing.
    \item \textbf{Roy Cerqueti} Conceptualization, Methodology, Supervision, Writing.
    \item \textbf{Walter Quattrociocchi} Conceptualization, Supervision, Writing.
\end{itemize}

\section*{Declaration of Competing Interest} The authors declare no competing interest.

\section{Data availability}
The data collection and analysis process are compliant with the terms and conditions imposed by Crowdtangle \cite{crowdtangle}. Therefore, the results described in this paper cannot be exploited to infer the identity of the accounts involved. CrowdTangle does not include paid ads unless those ads began as organic, non-paid posts that were subsequently “boosted” using Facebook’s advertising tools. It also does not include activity on private accounts, or posts made visible only to specific groups of followers. 

\section{SI Appendix Section}
\label{sec:sample:appendix}

\bibliography{bibliography}

\end{document}